# Simulations of radiation damage in spacecraft camera for ESA JUICE mission

Hualin Xiao, Wojtek Hajdas, Stephane Beauvivre, Daniel Kraehenbuehl, Ruth Ziethe and Nikhil Banerji

*Abstract*— The JUpiter ICy moons Explorer (JUICE) is an ESA interplanetary spacecraft being developed to perform detailed investigations of the Jupiter system and three of its icy moons: Europa, Callisto and Ganymede. The emphasis will be given on Ganymede as a small planetary body to be studied as a potential habitat. The spacecraft is set for launch in 2022 and would reach Jupiter in 2030. Two identical optical cameras are proposed for the mission to monitor the spacecraft and its surroundings. The sensors of the cameras need to be protected from hazardous radiation levels caused by extremely high fluxes of very energetic electrons. A precise model of the camera was developed to be used for intense Monte Carlo simulations performed to optimize the shielding and to determine the radiation damage during the mission. Simulations included determination of the total ionizing and non-ionizing doses in the sensors and crucial electronic components. This paper presents both simulation methods and results.

## I. INTRODUCTION

THE JUpiter ICy moons Explorer (JUICE) is an ESA interplanetary spacecraft being developed to perform detailed investigations of the Jupiter system and three of its icy moons: Europa, Callisto and Ganymede. The emphasis will be given on Ganymede as a small planetary body to be studied as a potential habitat. The spacecraft is set for launch in 2022 and would reach Jupiter in 2030. Two identical optical cameras (JMCs) are proposed for the mission to monitor the spacecraft and its surroundings. The sensors of the cameras need to be protected from hazardous radiation levels caused by extremely high fluxes of very energetic electrons. A precise model of the camera was developed to be used for intense Monte Carlo simulations with Geant4 performed to determine the camera performance including its radiation damage and image quality degradation during the mission. Computations were done for several optimized shielding concepts and relied on existing

Jupiter radiation models. They included determination of the total ionizing and non-ionizing doses in the sensors and crucial electronic components.

In the following chapters, we present the mass model, simulations and parts of results.

## II. SIMULATION PACKAGE

A complete Monte Carlo simulation package was built based on the Geant4 suite developed by CERN [1]. The QGSP physics model is used as the physics list for total ionizing dose calculation. In addition to QGSP physics list, the Geant4 Coulomb scattering process was enabled for non-ionizing dose simulations. Incident particle definitions used either General Particle Source toolkit (GPS) from Geant4. The geometry description of the camera was converted from a STEP file with a tool developed at PSI [2]. The left panel of Fig. 1 shows the view of the latest design of the camera with Geant4. In addition to JMC model a simplified model of the spacecraft was also introduced as presented in the right panel of Fig. 1. One can also see two small cameras located on the spacecraft.

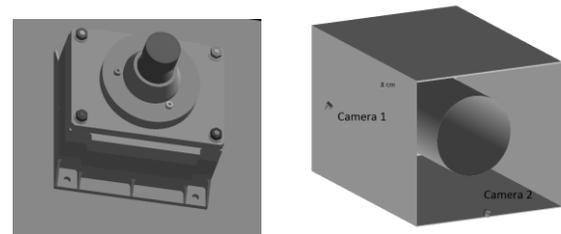

Fig. 1 Left - View of the latest design of JUICE mission camera with GEANT4. Right – View of the JUICE spacecraft mockup with two cameras from MicroCamera attached to its structure.

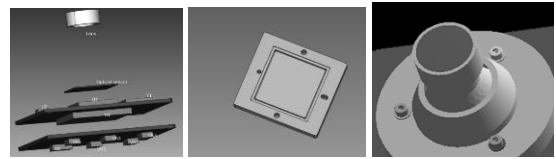

Fig. 2 Left - Visualization of all sensitive volumes. Center –JMC image sensor with is housing frame. Right - The 1$^{st}$ lens on top of the objective structure.

There were in total twelve sensitive volumes in the mass model: the image sensor, the first lens of the optical system as well as ten electronic chips models. All sensitive volumes are presented in Fig. 2. Ten dummy chips were placed in accordance to the geometry design seen in the Figure above (left). Their dimensions were provided in the STEP file and correspond directly to the real chips selected for the electronics of the micro-camera.

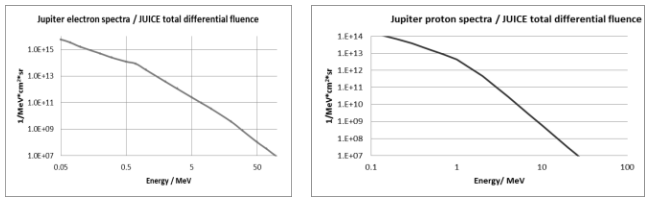

Fig 3. JUICE mission integrated electron fluence and proton fluence used for simulations.

Dose calculations were performed for both electrons and protons. The spectra used in the calculations are based on the JUPITER radiation model constructed for the JUICE mission purposes. Particles were generated on the surface of a sphere of 500 cm radius surrounding one of the cameras. The energy spectra of the primary particles are shown in Fig. 3. The directions of particles followed a cosine law and were limited to a cone of 3 degree. The beam spot was selected to be large enough to fully cover the whole camera and its surroundings. The particles were emitted from all directions. The total ionizing dose for various mission segments of the flight starting from the transfer to Europe (phase 1) to Europa flyby (Phase 2), high latitude pass to Callisto (phase 3), transfer to Ganymede (Phase 4) to the final insertion to Ganymede (Phase 5a) and Ganymede circular orbit one (Phase 5b) were simulated. Simulation runs were performed for each of two cameras separately.

### III. SHIELDING OPTIMIZATION

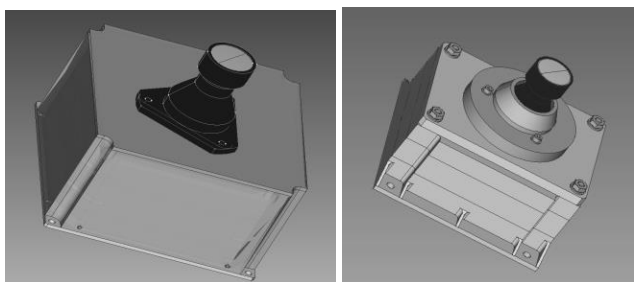

Fig. 4 JUICE camera before (left) and after (right) shielding optimization.

The left panel of Fig. 4 shows the first geometry design of camera. The lens housing material was tungsten. The material for the side shielding on the top and the bottom was aluminum.

The total ionizing total simulated for the image sensor for the whole mission was equal to 140 krad. The shielding was optimized in order to reduce the total ionizing dose. The main modifications were as follows:
- The top cover material was replaced with Titanium;
- A shielding ring was added in front of the top cover in order to reduce radiation from the top;
- The lens housing material was replaced with Titanium;
- The bottom and the side shielding was enhanced by increasing the thickness.

The right panel of Fig. 4 shows the view of the camera after shielding optimization. The total ionizing dose for the optimized model during the whole mission is about 40 krad. It is worthwhile to remark that the optimization was limited by the small mass budget of about 1 kg. Optimization of the shielding is still ongoing.

### IV. RESULTS FOR THE CAMERA AFTER SHIELDING OPTIMIZATION

#### A. Particles Crossing the Image Sensor

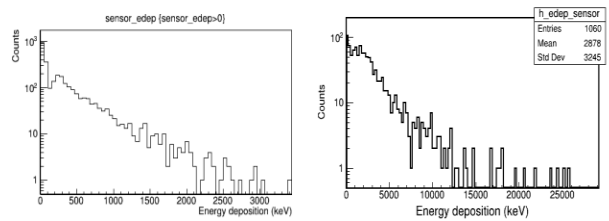

Fig. 5 Distribution of energies deposited in the image sensor by electrons and protons modeled for the Jupiter radiation environment.

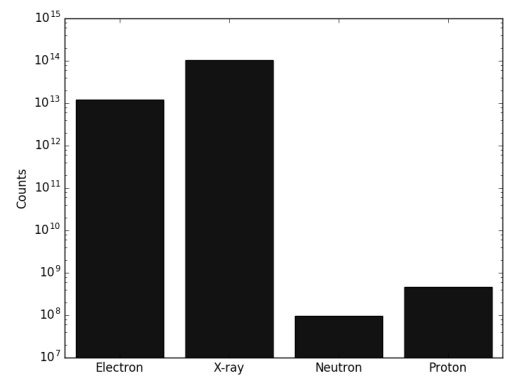

Fig. 6 Species entering the image sensor.

Fig. 5 show the energy depositions caused by primary electrons and protons. Fig. 6 shows distribution of particle species crossing the sensor surface. As can be seen the majority of the particles entering the sensor are X-rays produced via Bremsstrahlung and penetrated electrons. The contribution of primary protons in total ionizing dose is negligible compared to electrons.

#### B. Total Ionizing doses

The Total ionizing dose (TID) results for different mission segments as well as for the whole mission are shown in Table 1. As one can see the highest damage i.e. more than 50% of the total dose damage will occur only at the last 20% of the mission period during JUICE Ganymede phases 5a and 5b.

TABLE 1 SUMMARY OF DOSES FRACTIONS (IN KRAD) OVER ALL JUICE MISSION SEGMENTS IN VARIOUS CAMERA PARTS.

| Mission segments | Mission phases | | | | | | |
|---|---|---|---|---|---|---|---|
| | 1 | 2 | 3 | 4 | 5a | 5b | Total |
| Camera 1 image sensor | 8.4 | 6.0 | 3.0 | 6.9 | 11.7 | 2.8 | 38.8 |
| Camera 2 image sensor | 9.4 | 7.3 | 4.2.0 | 2.8 | 16.7 | 3.6 | 44.0 |
| Lens 1 | 551.0 | 188.0 | 438.0 | 1052.0 | 2133.0 | 749.0 | 5111.0 |
| Lens 1; 1st 1 mm | 3394.0 | 1180.0 | 2742.0 | 6483.0 | 13265.0 | 4586.0 | 31650.0 |
| Lens 1; 2nd 1 mm | 431.0 | 150.0 | 349.0 | 824.0 | 1686.0 | 583.0 | 4023.0 |
| Chip VR5 (worst case) | 20.0 | 17.3 | 17.9 | 27.6 | 94.0 | 10.4 | 187.2 |

*C. Non-ionizing (displacement damage) Dose Calculation*

For all the simulations runs described above the particle momentum direction, its kinetic energy, energy loss as well as its type were recorded and stored in the output files if it entered the sensitive volumes. Fig. 6 and Fig. 7 show examples of the energy spectra of the recorded particles for two runs with initial particles being either electrons or protons respectively. It should be noted that also the particles which did not deposit energies in the sensor were stored. A new simulation package was prepared for displacement calculations. This package includes physics of single Coulomb scattering process as well as nuclear reactions. The mass model of this package consists only of the sensitive volumes. The previously recorded articles entering the sensitive volumes are used as the primary particles.

Calculations of NIEL doses were based on the following equation [3]:

$$-\left(\frac{dE}{dx}\right)^{NIEL}_{nucl} = n_A \int_{T_d}^{T_{max}} T\, L(T)\, \frac{d\sigma^{WM}(T)}{dT}\, dT,$$

where $n_A$ is the number of atoms per cm$^3$ in the absorber; $T$ and $T_{max}$ the recoil kinetic energy and the maximum energy transferred to the recoil nucleus, respectively; $L(T)$ is the Lindhard partition function (see Ref. [3]) ; $T_d$ is threshold energy for displacement and $d\sigma/dt$ is the differential cross section for elastic Coulomb scattering. The product $T\,L$ is also called the damage energy. In the case of silicon, $T_d$ is equal to 21 eV.

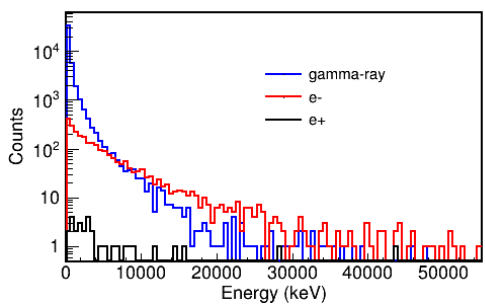

Fig. 6 Energy spectra of particles entering the image sensor recorded during electron simulation runs.

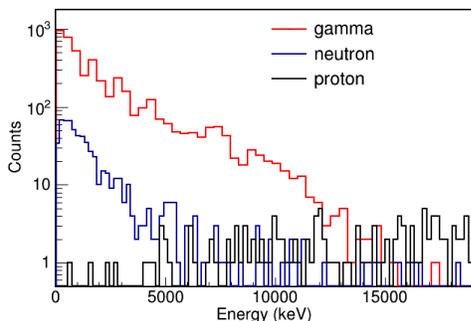

Fig. 7 Energy spectra of particles entering the image sensor recorded during proton simulation runs.

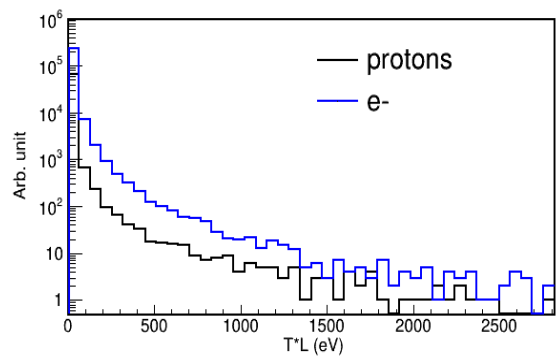

Fig. 8 Spectra of the damage energy $LT$ in the image sensor.

Fig. 8 shows examples of the damage energy spectra for proton and electron runs. The total dose values were subsequently calculated by summing over the damage energies. In the same way as for the TID calculations the total NIEL dose deposited in the sensor during the mission is obtained by scaling the simulated dose values to the fluence of the whole mission. The average non-ionizing stopping power values, NIEL doses for the image sensor and the chip VR5 are shown in Table 2. The 50 MeV proton equivalent fluences are also presented in the table.

TABLE2. NIEL STOPPING POWER VALUES FOR THE IMAGE SENSOR AND THE CHIP VR5

| sensor/chip | DD stopping power (MeV/cm) | DD mass stopping power (MeV cm2/g) | Total NIEL dose (MeV/g) | 50 MeV proton equivalent fluence (p/cm2) |
|---|---|---|---|---|
| Image sensor | 2.8e-3 | 1.2e-3 | 6.64E7 | 1.51E10 |
| Chip VR5 | 1.27e-2 | 5.46-3 | 7.92E8 | 1.79E11 |

## V. CONCLUSION

Two micro-cameras are proposed for ESA JUICE mission. A precise model of the camera was developed to be used for intense Monte Carlo simulations performed to optimize the shielding and to determine the radiation damage during the mission. Simulations included determination of the total ionizing and non-ionizing doses in the sensors and crucial electronic components. The TID calculated for the image sensor for the whole mission is about 40 krad and the TIDs for the crucial chips are between 100 krad and 200 krad. The NIEL doses simulated for the image sensor and the chip VR5 are equal to 6.64E7 MeV/g and 7.92E8 MeV/g respectively.

REFERENCES

[1] S. Agostinelli, et al. Geant4 a simulation toolkit Nucl. Instrum. Methods Phys. Res. A, 506 (2003), p. 250
[2] CAD file to Geant4 GDML converter: http://polar.psi.ch/cadmc/
[3] C. Leroy and P.G. Rancoita (2016), Principles of Radiation Interaction in Matter and Detection - 4th Edition -, World Scientific. Singapore.